# Virtual academic conferencing: a scoping review of 1984–2021 literature. Novel modalities vs. long-standing challenges in scholarly communication

Agnieszka Olechnicka[1*], Adam Ploszaj[1], Ewa Zegler-Poleska[1]
[1] Science Studies Laboratory, Centre for European Regional and Local Studies EUROREG, University of Warsaw, Warsaw, Poland
*Corresponding author. Email: a.olechnicka@uw.edu.pl

**Abstract**
Objective. This study reviews the literature on virtual academic conferences, which have gained significant attention due to the COVID-19 pandemic.
Design. We conducted a scoping review, analyzing 147 documents available up to October 5th, 2021. We categorized this literature, identified main themes, examined theoretical approaches, evaluated empirical findings, and synthesized the advantages and disadvantages of virtual academic conferences.
Results. We find that the existing literature on virtual academic conferences is mainly descriptive and lacks a solid theoretical framework for studying the phenomenon. Despite the rapid growth of the literature documenting and discussing virtual conferencing induced by the pandemic, the understanding of the phenomenon is limited.
Conclusions. We provide recommendations for future research on academic virtual conferences: their impact on research productivity, quality, and collaboration; relations to social, economic, and geopolitical inequalities in science; and their environmental aspects. We stress the need for further research encompassing the development of a theoretical framework that will guide empirical studies.

## 1. Introduction

As gatherings of professionals, conferences have been the object of interest in various areas of research, including event management, management studies, higher education, and regional studies. Conferences are temporary clusters of interactions or hotspots of intense knowledge exchange (Maskell et al., 2006). Moreover, conferences are venues for various activities ranging from information exchanges to the enactment of technological possibilities (Garud, 2008). More recent developments in conference studies consider them as key sites of knowledge creation, public performance, legitimation, and protest (Craggs & Mahony, 2014). The return on investment from conferencing has also received scrutiny (Edelheim et al., 2018).

Academic conferences are a crucial element of scientific work and scholarly communication. Their essential role as a meeting space became apparent during the COVID-19 pandemic and the ensuing travel restrictions. However, multiple authors have noticed that despite the long tradition and essential role of conferences in science, academic meetings have been largely neglected as a subject of research (Hansen & Budtz Pedersen, 2018) even in areas studying science. For example, a typology of academic events has been only recently provided (Hansen et al., 2020).

Among the few studies exploring academic events a prominent topic is their epistemic and social role. Knorr-Cetina (1995) considers scientific meetings in high-energy physics as places of consensus formation. Similarly, Craggs and Mahony (2014) analyze conferences as places producing epistemic communities. Along this line, Hansen and Budtz Pedersen (2020) postulate conceptualizing and evaluating academic events as open marketplaces that facilitate the conversion of credibility.

Primarily concerned with the social functions of academic conferences, Gross and Fleming (2011) contribute to their understanding as “key sites for the social orchestration of academic knowledge and the intrusion of sociality into forms of social knowledge production.” Overall functions of conferences have been conceptualized by Jacobs and McFarlane (2005), who integrated insights from science studies and sociocultural theory. According to these authors, conferences are formalized arenas where procedures of science are examined and held accountable; they provide space for managing interpretive flexibility and for operating closure mechanisms. On the other hand, conferences function as communities of practice and places of knowledge-building.

Similar to the lack of research on academic conferences in general, virtual academic conferences (VCs) have not been extensively studied until the COVID-19 pandemic, when they became the only feasible mode to gather scholars. For this reason, the topic of academic VC has been attracting growing attention in many disciplines forced to organize meetings in virtual mode. This growth has also resulted in a stronger interest in VC as an object of study. While VCs as professional development events have already been thoroughly described over a decade ago by Anderson (2010), even recent works argue that virtual conferences have not received much attention because of their newness (Fang & Daniel, 2021). The extant reviews on academic conferences and their virtual modes are limited (González-Santos & Dimond, 2015; Sá et al., 2019).

Parallel to the recent proliferation of virtual conferences, the amount of literature documenting and discussing these events has also rapidly grown. This unprecedented growth demands closer examination. Our literature review aims to capture the characteristics of this burgeoning body of works and provide an overview of potential research niches. In particular, we aimed to answer the following research questions:

(1) How are virtual conferences defined?

(2) What are the main themes discussed in the literature on VCs?

(3) What theoretical approaches to VCs are employed?

(4) What is the state of empirical research on VCs?

(5) What are the advantages and disadvantages of VCs?

We conduct a scoping review with elements of mapping review, providing a narrative description of the current state of research on VC, based on a comprehensive search (Grant & Booth, 2009). We categorize the existing literature, assessing its quantity and quality. We do not perform a formal quality assessment but focus on identifying gaps in the literature where more research is needed.

## 2. Data and methods

We searched Scopus, Web of Science databases, and arXiv and OSF preprints repositories with combinations of keywords "conference" and "virtual" and their synonyms and related terms, using truncation and Boolean operators. As we focused on academic conferences, we refined the search to exclude records concerning other types of meetings, such as business events and remote healthcare, telemedicine, or teaching seminars. Our inclusion criterion was that the academic virtual conferences must be the main topic of the publication with no restrictions to the type of publication. As exclusion criteria, we defined: (1) educational virtual events focused on pupils or undergraduate students, (2) remote healthcare and telemedicine meetings, (3) business, and industry online events being the main topic, (4) non-English language of publication. Otherwise, the investigation included all indexed types of publications with no restriction on publication date. The final query used in this study was written as follows (this example uses SCOPUS syntax):

(TITLE-ABS-KEY ( "videoconferenc*" OR "webconferenc*" OR "virtual conferenc*" OR "video conferenc*" OR "web conferenc*" OR "on-line conferenc*" OR "online conferenc*" OR "digital conferenc*" OR "remote conferenc*" OR "e-conferenc*" OR "hybrid conferenc*" OR "virtual seminar*" OR "video seminar*" OR "web seminar*" OR "on-line seminar*" OR "online seminar*" OR "digital seminar*" OR "remote seminar*" OR "e-seminar*" OR "hybrid seminar*" OR "virtual congress*" OR "video congress*" OR "web congress*" OR "on-line congress*" OR "online congress*" OR "digital congress*" OR "remote congress*" OR "e-congress*" OR "hybrid congress*" OR "virtual scientific event*" OR "video scientific event*" OR "on-line scientific event*" OR "online scientific event*" OR "hybrid scientific event*" ) AND TITLE-ABS-KEY OR ( "scientific" OR "academic*" ) AND NOT TITLE-ABS-KEY ( "teach*" OR "student*" OR "patient*" OR "telemed*" OR "educat*" OR "industry" OR "care hom*" OR "telehealth" OR "physician*" OR "interview*" ))

Following the scoping review methodology, we conducted multiple structured searches (Grant & Booth, 2009). Through an iterative process of testing and refining the search combined with the snowballing procedure (Gough et al., 2017), we identified 296 documents for further selection. Additionally, we screened the included articles' backward (references) and forward citation reference lists to identify relevant complementary literature sources. Then we sequentially examined the title and abstract of each publication as well as the content when needed. As a result of this selection, we obtained 92 documents. An update of this process conducted in October 2021 resulted in further 55 documents.

After these steps, our final set included 147 documents and covered the literature available up to October 5th, 2021. We have offered the entire set of publication metadata online as a Zotero Group "Virtual Academic Conferencing 1984–2021" (https://www.zotero.org/groups/4832623/virtual_academic_conferencing_1984-2021).

We collected full texts of the publications and imported PDF files into MaxQDA 2020 (VERBI Software, Berlin, Germany) to conduct a content analysis, combining deductive and inductive approaches. For that purpose, we created a coding framework. We predefined the set of codes accordingly to our research questions. We annotated each of the codes with a detailed memo for common understanding within the coding team. After the initial coding phase, each code was further re-coded inductively, and several subcodes were applied to analyze the retrieved fragments.

## 3. Results

### 3.1. Types of publications and venues

To compare the pre- and during-pandemic trends, we divide the collected 147 publications into two groups: (1) published until 2019 inclusive, i.e., before the pandemic, and 2) published in 2020 and 2021, during the pandemic. The former group comprises 35 items published in 1984–2019, the latter 112 published in 2020 and 2021. The latter group includes eight papers with publication date 2020, but apparently submitted early enough not to concern the pandemic (Abbott, 2020; Arnal et al., 2020; Black et al., 2020; Ekstrom et al., 2020; Fellermann et al., 2020; Guerra Amorim & Tucci, 2020; Le et al., 2020a; Spilker et al., 2020).

The majority of documents analyzed (121; 82%) were published in academic journals; the remaining were: 10 conference papers, 8 preprints, 3 blog posts (BLOG@CACM, LSE Impact Blog, eLife Labs), 2 magazine articles ("ACM Interactions," "The Bulletin of the Ecological Society of America"), 1 book chapter, 1 report, and 1 working paper. Types of publications were verified and corrected manually, as several records appeared miscategorized after the Zotero import.

Most venues were represented once, with several exceptions of journals: *Nature* (7), *PLOS Computational Biology* (4), *Elementa: Science of the Anthropocene* (3), Educational Technology and Society (3), and *eLife* (3). Furthermore, six of eight preprints appeared in the "Computer Science" section in arXiv. Note that despite the prevalence and variety of journals represented in the collection, most items are editorial materials and similar non-refereed publications. Among 91 items for which the journal specified the type or section, only 20 were listed as research papers, whereas the majority 71 items were listed under various headings such as commentaries, editorials, news, etc. For example, three *Nature* publications were listed as news, one as a comment, and three in sections providing advice on scientific careers. Similarly, all four *PLOS Computational Biology* publications were editorials in the "Ten simple rules" guide series.

### 3.2. Temporal dimension – before and after COVID-19

Although the first publications about virtual conferences appeared as early as the 1980s, it was not until the COVID-19 pandemic that a huge wave of interest in the topic was generated (see Fig 1). 2020 represents an important caesura in publications on virtual conferences in the academy.

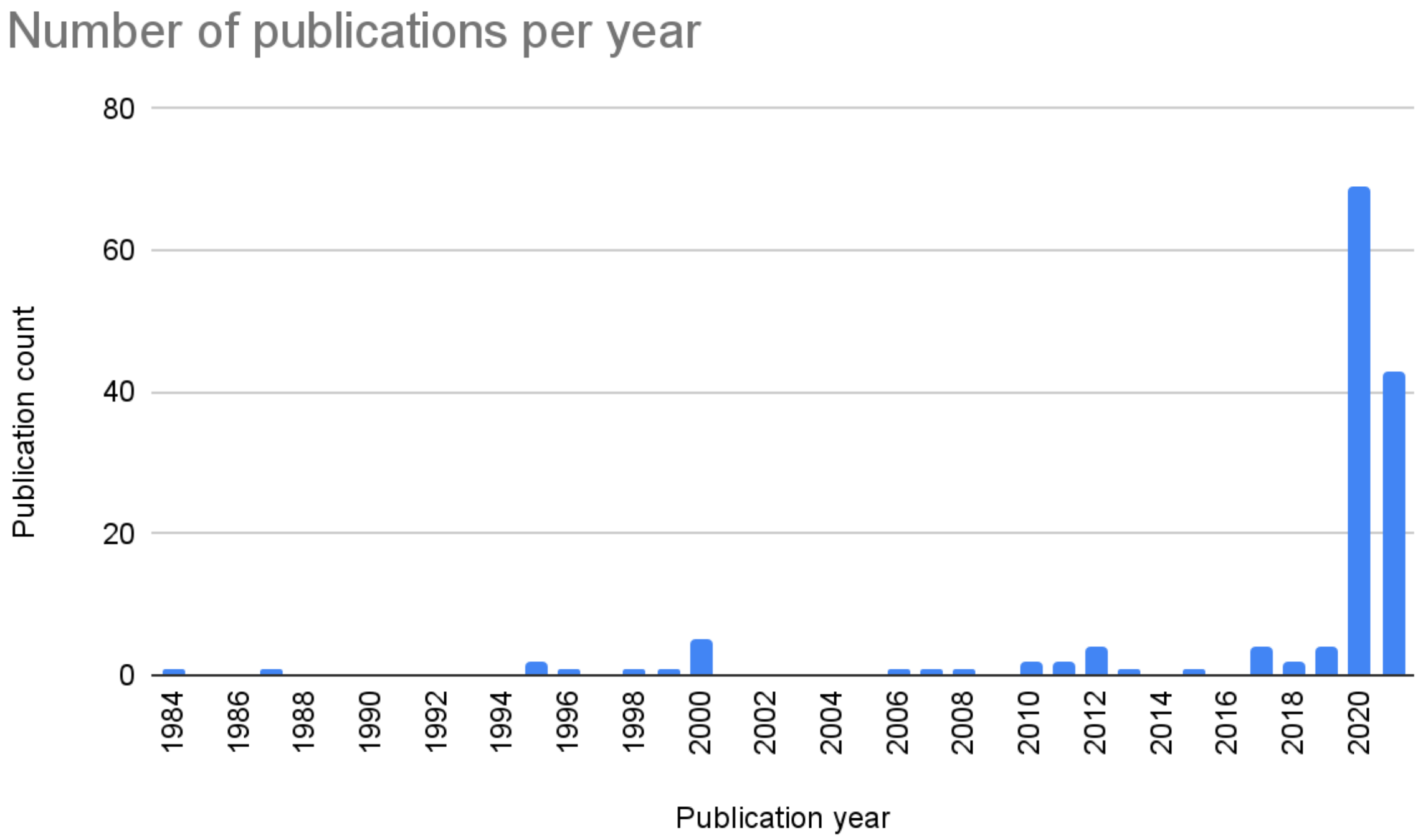

Fig 1. Publications included in the review per year

The pre-pandemic part of the collection (1984-2019) includes 35 documents: 29 journal publications, three conference papers, one report, one book chapter, and one magazine article. Regarding the subject matter of publication venues, we categorized these documents into several groups. Most (22; 63%) were published in venues representing a range of social sciences and computer sciences. More than half (19) of the documents appeared in venues representing education and computer sciences as well as the nexus of these areas, such as educational technology (a report by Green (1998) also belongs to this group). Three were published in library and information science journals (Baird & Borer, 1987; O'Haver, 1995; Peuler & McCallister, 2019); one review article was published in a journal focused on hospitality and tourism (Sox et al., 2017). Among the remaining 14 documents, four (11% of all) appeared in medical journals, including an article in the *Journal of the American Medical Informatics Association* (Lecueder & Manyari, 2000). Two documents represented chemistry, including one in the *Journal of Chemical Information and Computer Sciences* (Bachrach, 1995), and one was published in *PLOS Computational Biology* (Gichora et al., 2010). Again, the nexus of the subject matter and informational technology is noticeable. Three documents were published in venues focused on environmental issues: *Conservation Biology* (Fraser et al., 2017), *The Bulletin of the Ecological Society of America* (Hampton et al., 2017), and *Elementa: Science of the Anthropocene* (Parncutt & Seither-Preisler, 2019). Multidisciplinary *Nature* and *Science* were represented once each.

The post-COVID-19 (2020–2021) group comprises 112 documents, the vast majority of which (92) were published in journals. The remaining 20 items include eight preprints, seven conference papers, three blog posts, one magazine article, and one working paper. Concerning subject areas, most were published in medical journals, with 27 items (24%). The next largest groups are the social sciences (19), particularly education, including educational technology (6); biological sciences – 18 (16%); and computer sciences – 17 (15%). Other notable, though smaller, groups were environmental and ecology – 8 items and multidisciplinary venues – 7.

The most noticeable characteristic of the analyzed collection is the prevalence of social and computer sciences venues in the pre-pandemic period (63%), which later changes in favor of medical and biological sciences (40%). In the pandemic period, social and computer sciences' share is still substantial, but reduced twice (32%), while medical increases twofold (from 11% to 24%) and expands to various branches; biological increases even more (from 3% to 16%). At the same time only four documents appeared in journals dedicated to science studies or science of science: two in *Scientometrics* (Falk & Hagsten, 2021b; Mubin et al., 2021) one in the *Journal of Responsible Innovation* (Braun et al., 2020) and one in the *British Journal for the History of Science* (Robinson et al., 2020). However, the last document is a case study of the conference "British Society for the History of Science."

### 3.3. VC definition, evolution, and uptake

In that one of our goals was to provide an overview of virtual conferences, we did not establish a prior definition of VC. Thus, our inclusive search strategy resulted in a collection of literature that covers a range of interpretations of the notion of virtual conference. Some of the collected works notice a difficulty in defining what a virtual conference is (Wilkinson & Hemby, 2000) and the need to define terms such as 'online conferencing' (Ball, 2000). Upon closer examination, it appears that only about a dozen publications provide an explicit definition of VC, of which only a few are more detailed (T. Anderson, 1996; Sá et al., 2019; Thatcher et al., 2011). However, in many other works, the implicit understanding of VC can be gleaned from the description of the event reported. Understandably, defining VC is more elaborated in earlier publications, when virtual events were a novelty; recent publications tend to address it in a cursory manner.

VC evolution concurrent to technological developments we discerned in our collection is consistent with the trends reported in (L. Anderson, 2010). The two works from the 1980s (Baird & Borer, 1987; Tombaugh, 1984) discuss early beginnings of text-based computer conferencing as the then-emerging form of computer-mediated communication. Next, papers published in the 1990s and 2000s provide examples of introducing VC in various areas, such as chemistry education (Bachrach, 1995; O'Haver, 1995), distance education (T. Anderson, 1996), psychiatry (Batra et al., 1999), political science (Ball, 2000), cardiology (Lecueder & Manyari, 2000), and ergonomics (Thatcher, 2006). Additionally, a guide by Green (1998) is based on experience in organizing conferences in the area of learning technologies.

VC are usually characterized in comparison to their face-to-face (F2F) predecessors. This strategy is not particularly fruitful, as F2F academic conferences are still an understudied topic. However, several publications attempt to utilize the comparison to gain more insight. For example, Jones (2000) offers a taxonomy of comparative factors related to both types of conferences, and Schwarz et al. (2020a) propose to systematize academic interactions and the corresponding digital formats. Meanwhile, (Sá et al., 2019) review the advantages, limitations, and potentials of F2F and VC. Another issue is distinguishing VC from other types of mediated communication (Sá et al., 2019; Schwarz et al., 2020a; Thatcher et al., 2011; L. J. L. Veldhuizen et al., 2020).

A comprehensive definition provided by T. Anderson (1996) characterizes VC as a structured, time-bound professional development activity using technologies to support communication and interaction between geographically distributed participants, taking place synchronously, asynchronously, or both. Generally, the literature appears to share this understanding of VC and some discrepancies arise due to the three major dimensions of virtual events: technology, time, and location.

A typology proposed by Fraser et al. (2017) consists of four virtual conference models: (1) pure virtual, in which participants interact online from external locations linked by a conference virtual network; (2) one hub and node, in which a central hub hosts and streams the conference to nodes – smaller external venues; (3) multihub and node, involving multiple international hubs within a similar time zone; (4) multilateral hub and node, comprising multiple hubs across multiple time zones.

### 3.4. Topics and threads

Thematically, the analyzed collection of articles is fairly homogeneous. Three-fourths of the papers address the issue of virtual conferences in a general way, covering many aspects in a single article, attempting to approach the phenomenon holistically and often with a practical guidebook attitude. More than half (53.1%) of the items can be characterized as describing lessons learned based on the experience of conference organizers. At the same time, most of these articles (44.9% of the total) are based on experiences from one conference or one conference series (e.g., a cyclical conference moved to virtual during the pandemic). Only a small fraction of lessons learned papers are based on experiences from different virtual events (Bottanelli et al., 2020; Carr & Ludvigsen, 2017; Forrest et al., 2020; Green, 1998; Hampton et al., 2017; M. L. W. Jones, 2000; Meyer et al., 2021; Mikhridinova et al., 2021; Rubinger et al., 2020; Wang et al., 2020; Weissgerber et al., 2020; Wilkinson & Hemby, 2000). Lessons-learned articles, as a rule, are often clearly advisory in nature and are cross-cutting, addressing various aspects of the organization and operation of virtual conferences. These articles provide tips and "simple rules" for organizers of future virtual conferences, and less frequently, for participants.

One in four articles (25.9%) can be characterized as presenting general thoughts on VC and are not based on experience from any specific event or events or any systematic analysis. These articles have the characteristics of news, essays, or editorials – they openly have no scientific pretensions. Often such articles have a broad spectrum of addressed topics, such as the lessons-learned group discussed above. General articles frequently discuss the challenges of holding a virtual conference, point out advantages and disadvantages, as well as divagate on future directions for online meetings or scholarly meetings in general.

A small part of the analyzed collection of papers has a clear thematic focus, which, by the way, often correlates with the more scientific nature of these articles (but not always). One theme stands out in this group, i.e., the environmental aspects of virtual conferences, sometimes in comparison with traditional conferences, sometimes as a proposal of solutions to be implemented. The environmental thread, especially in the context of the carbon footprint of virtual conferences, appears more often as one of the enumerated advantages of virtual conferences, but only a few articles explicitly focus on this issue (Caravaggi et al., 2021; Coroama et al., 2012; Ekstrom et al., 2020; Fellermann et al., 2020; Jäckle, 2021; Klöwer et al., 2020; Lester, 2007; Parncutt & Seither-Preisler, 2019; Rubinger et al., 2020).

Another distinct theme concerns the study of engagement and the social functions of conference attendance. As with the thread on the carbon footprint of conferences, the social interaction thread is very often mentioned as an advantage or disadvantage of a particular type of conference, but only a few papers focus on this issue (Caravaggi et al., 2021; Coroama et al., 2012; Ekstrom et al., 2020; Fellermann et al., 2020; Jäckle, 2021; Klöwer et al., 2020; Lester, 2007; Parncutt & Seither-Preisler, n.d., 2019; Rubinger et al., 2020).

In addition, in the case of few articles, one can speak of specific narrow topics concerning virtual conferences. These are articles describing the reactions of conference organizers to the pandemic, presenting statistics on the cancellation of conferences, the transition to a virtual or hybrid form, or analyzing the determinants of these decisions, or show what the consequences are of canceling a conference (Falk & Hagsten, 2021b; Lessing et al., 2020; Weissgerber et al., 2020), pricing of virtual conferences (Falk & Hagsten, 2021a), discuss questionnaire methodology to assess the participants experience of virtual conferences (Hohlfeld et al., 2021), or present specific technical setups for VC, usually focused on enhanced engagement and interaction (Park et al., 2020; B. Rogers et al., 2018; Song et al., n.d.; Wang et al., 2020; Wu et al., 2021).

**3.5. Data and methods in reviewed publications**

A characteristic feature of the analyzed publications is their relatively lightweight methodological sophistication. To start with, slightly more than half (55.8%) of the publications contain any explicit reference to the method or data used. For the rest of the set, the nature of the text indicates that the work is based on the authors' personal experience, mainly as organizers of virtual conferences. However, usually the methodology is not explicitly indicated but implicitly contained in the text. To some extent, this ambiguity can be explained by the fact that many papers in our collection are published as opinion pieces, notes, or editorials.

Surveys were by far the most widespread way of collecting data in the analyzed set of articles, with 36% of articles based on survey data. The surveys are almost exclusively post-conference evaluation surveys conducted after a single event, based on a non-random sampling. Sample sizes are somewhat on the smaller side. Out of 53 survey-based articles, 28.3% employed sample sizes of less than 100 (the smallest sample size is as low as 22). Furthermore, proper reporting on sample size is often missing. For 22.6% of survey-based articles there is no sample size information. In addition, in some articles, sample sizes are reported approximately ("about 200" or "nearly two hundred responses"). The largest sample size is 2120 (Gao et al., 2020), however, it is a non-random convenience sample survey in which participation has been solicited via social media platforms. Target populations are rarely described in an explicit way, sometimes being implicit, that is, equal to the participants of a given event. Consequently, response rates are rarely reported.

More complex survey designs are rare. Pre- and post-surveys are used to compare the expectations and experiences of participants of conferences that have been moved virtual (McDowell et al., 2020; Misa et al., 2020). Repeated surveys are used to trace changes that occur during multi-day events (Dunn et al., 2021; Mikhridinova et al., 2021). Other authors compare post-surveys of in-person pre-pandemic conferences with their pandemic virtual editions (Stamelou et al., 2021; Terhune et al., 2020). A few articles are based on surveys collected for more than one event (L. J. Veldhuizen et al., 2020) or for different editions of a periodical event (Ho et al., 2011; Thatcher et al., 2011). One of the other articles from a pre-pandemic wave of virtual conferences papers stands out considering survey design. (Wilkinson & Hemby, 2000) compare opinions of randomly sampled participants of two virtual conferences and two control groups composed of non-participants randomly selected from the population of members of associations organizing those conferences (total sample size: 267).

A distinct group of papers uses registration and organizational data on one or more conferences. For single conference papers, this kind of data is used primarily to present demographic information on conference participants. For multiple conference papers, registration and organizational data uses are more creative. Based on data on 587 conferences, the probability of conferences switching to virtual during the pandemic has been modeled (Falk & Hagsten, 2021b). The same authors used a smaller data set of 76 virtual conferences to analyze the variation in conference fees (Falk & Hagsten, 2021a). However, typically this kind of data is used to present changes, for instance, the number of participants by country, in an annually held conference(s) (Harabor & Vallati, 2020; Sarabipour, 2020; Stamelou et al., 2021; Thatcher et al., 2011) or different responses to pandemic, for instance,, postponed, canceled, etc. – in a larger sample of conferences (Ha et al., 2021; Mubin et al., 2021).

Roughly one in ten articles use data automatically collected from servers. This kind of data is typically employed to report usage statistics of conference websites, virtual conferencing software, or related communication and social media channels (e.g., (Le et al., 2020a). The potential of this type of data seems to be underused in the analyzed literature. Not only are they relatively rarely used, and if they are, the use is usually purely descriptive. The work of (Wu et al., 2021) stands out against this background as it creatively combines the use of automatically obtained data with survey data. The other interesting approach is presented by (Reinhard et al., 2020), who count the number of questions addressed to individual panelists based on the video recordings of conference panels.

The use of quantitative data in the collected collection of articles is usually very basic, limited to giving counts or percentages. Only a few articles (about 5 % of the set) use more serious methods of analysis, such as significance testing or regression analysis (Dunn et al., 2021; Falk & Hagsten, 2021b; Gao et al., 2020; Hameed et al., 2021; Terhune et al., 2020; Wilkinson & Hemby, 2000; Wu et al., 2021).

Qualitative research methods are also rarely used adequately in the analyzed set of articles. Only a few articles are based on formal interviews (Ho et al., 2011; Jauhiainen, 2021; Kirchner & Nordin Forsberg, 2021; Pedaste & Kasemets, 2021), while a few others refer to "informal interviews" or "unsolicited feedback from participants." Similarly, specific methodological approaches to the analysis of qualitative data are rarely indicated. Among them individual articles refer to participant observation (Jauhiainen, 2021; G. Sharma & Schroeder, 2013), autoethnography (Black et al., 2020) and abductive content analysis (Pedaste & Kasemets, 2021). Also, a few articles use mixed methods, some of them referring explicitly to this approach (Ho et al., 2011), and others use mixed methods without directly invoking the term (e.g., Jauhiainen, 2021).

Five papers use systematic or scoping literature review as their main method. As many as three of them belong to the group of pre-pandemic articles (Sá et al., 2019; Sox et al., 2017; Spilker et al., 2020). The remaining two review articles published in the pandemic wave of papers focus on extracting good practices from the literature (Lortie, 2020; Rubinger et al., 2020).

**3.6. Theoretical frameworks**

Within our sample, just twenty papers relate to the existing theories, concepts or theoretical approaches. In general, the authors' strategies are characterized by a lack of depth in applying the theories. Four inspiring examples which investigate virtual conferences from various theoretical perspectives are worth discussing in more detail (Carr & Ludvigsen, 2017; Spilker et al., 2020; Thatcher et al., 2011; Wu et al., 2021).

Thatcher et al. (2011) analyzed the CybErg conference series using the virtual settlement theory framework (Q. Jones, 1997). The authors focused their investigation on four conditions required to build an online community: a common-public space allowing interaction of all community members, a level of sustainability and stability of community membership, a variety of contributors in terms of field of interest, biographical indicators etc., and a level of interactivity among community members. The paper's final part discusses the findings on the background of the four-stages online community evolution approach proposed by Gongla and Rizzuto (2001), namely: a building stage, an engaged stage, an active stage, and an adaptive stage of the online community (Thatcher et al., 2011).

On the other hand, the paper by Wu et al. (2021) was grounded in the psychological concept of self-efficacy in which four potential sources of self-efficacy development: mastery experiences, vicarious experiences, verbal persuasion, and emotional and physiological states were distinguished (Bandura, 1993). The authors investigated virtual conference attendees' behavior and psychological conditions by studying the mastery experience, competence, and engagement of 150 participants of the Taiwan e-Learning Forum in 2020. Based on the collected information, the authors proved the correlation between mastery experience and competence in virtual conferences and the level of engagement in online academic meetings (Wu et al., 2021).

Carr and Ludvigsen (2017) offered the analytical stance known as cultural historical activity theory (CHAT), which is formulated around five ideas: object-oriented activity and mediation by tools and signs; historicity; multivoicedness of activity; contradictions as a source of change; and a zone of expansive learning. The CHAT framework explains drivers for change in socio-technical systems based on the analysis of tensions, disturbances, and contradictions existing in the given system. The authors assumed VC to be complex socio-technical systems and analyzed the multifaceted interactions among VC elements to redesign future scientific online events. The empirical portion of the paper was based on a content analysis of 256 statements by participants of a series of online conferences organized by the Centre for Educational Technology at the University of Cape Town (Carr & Ludvigsen, 2017).

The systematic literature review on the learning effects and knowledge-building of technology-enhanced conferences (Spilker et al., 2020) is the specific example of theory application to the VC paper. The study's broader background relates to continuing professional development of academics in communities of practice or social networks. The authors incorporated the Value Creation Framework (VCF) (Wenger et al., 2011) to present the results of conference attendance on the proliferation of digital professional competencies of academics. The VCF allows assessing the value creation in communities of practice and social networks that participants and organizers of the conferences embody. The paper is structured accordingly to the five cycles of the VCF that correspond with five types of value created during technology-enhanced conferences: immediate value refers to participation in the networked learning activity, potential value refers to the knowledge capital that the network generates, applied value refers to the transformation of practices, realized value refers to the application of knowledge capital and its impact, and reframing value refers to the redefinition of strategies and discounts at both individual and institutional levels.

The above examples confirm no clear tendency to focus on particular theories in the papers analyzing the VCs. The authors' approaches vary significantly. We can, however, distinguish five theoretical aspects deployed in the selected literature to investigate issues of VCs: (1) developing and maintaining virtual communities, (2) process of communication within networks, (3) innovation process, (4) functioning of the global system of knowledge production, (5) technological development.

The first group embraces the theories related to **developing and maintaining scholarly communities** and their ability to create added value in terms of learning and production of knowledge, practices, and standards. The previously discussed concepts of virtual settlement theory (Thatcher et al., 2011), CHAT (Carr & Ludvigsen, 2017) and VCF in a community of practice (Spilker et al., 2020), we can supplement with the concept of community of inquiry adopted to education field (T. Anderson, 1996).

According to this concept, online academic conferences are perceived as virtual communities, contributing to permanent professional development during and beyond the event moment. Virtual conferences provide opportunities to actively integrate participants and create a community of learners (T. Anderson, 1996) or spaces of collective learning (Sá et al., 2019). VCs are seen as a potential arena for elected tribes formation in the context of the professional collaboration of scholars (Reinhard et al., 2020).

The second group consists of concepts related to **a communication process within networks**. Tombaugh (1984) evaluated one of the earliest examples of VCs (in the form of an e-mail exchange) as a formula of computer-mediated scholarly communication from a social psychology perspective. Tombaugh employed equity theory, where an equitable relationship exists when all participants are receiving relatively equal and fair rewards from the relationship (Hatfield et al., 1978). In particular, he used equity theory to prove that it is yet impossible to achieve an equitable relationship in the case of computer-based communication, with one of the reasons being low mutual trust resulting from a deficiency of nonverbal and paraverbal cues.

The geographical and physical proximity concept invokes similar arguments recalled in several papers. Remote participation disrupts professional network creation, development, sharing, and the application of knowledge, as well as degrades politeness in communication (Jauhiainen, 2021; Niner & Wassermann, 2021; G. Sharma & Schroeder, 2013). Tuckman's group development theory (Tuckman & Jensen, 1977) might also fall into this category. Moreover Chou et al. (2012) propose guidelines facilitating successful VCs according to the four stages of group development theory: formation, setting ground rules, managing conflict, and enhancing group performance.

The third group incorporates the **notion of innovation** or connotes the peculiarities of innovation processes. First, some of the proposed VC definitions explicitly refer to Christensen's types of innovation from business theory (Christensen, 1997). The VCs as disruptive or low-end innovations are considered in opposition to F2F conferences. All papers drawing attention to the trust building in the networks (see group 2) could be counted in this group as well (L. Anderson, 2010; D. Sharma, 2021). Moreover, a review of hospitality and tourism papers on VCs used diffusion of innovation theory (E. M. Rogers, 2003) as an analytical framework (Sox et al., 2017). Jauhiainen (2021) explored the outcomes of the SHIFT conference organized via a 3D virtual platform, proposing a model of innovation development, including contexts of physical, virtual, and extended realities and interactions between humans and machines. Meanwhile, Braun et al. (2020) proposed using the responsible research and innovation approach (RRI-heuristic), applying the anticipation, inclusion, reflection and responsiveness (AIRR) approach to identify and reflect on the dilemmas involved in the virtualization of the research.

The fourth group of concepts relates to the **global system of knowledge production characteristics**. Goebel et al. (2020) refer to Latour's work (1987) on the impact of the scholarly migration on the knowledge production to deliberate opportunities and challenges given by the virtualization of conferences. Furthermore, Hansen and Pedersen (2018) acknowledge that the concept of credibility cycles (Latour & Woolgar, 1986) provides a helpful framework for studying how attending events influence the individual scholar's scientific potential. Shelley-Egan uses Urry's (Urry, 2003) concept of "meetingness" to discuss reduced mobility of scholars due to the pandemic and its consequences on the knowledge production (Shelley-Egan, 2020).

Several authors underline inequalities in opportunities and involvement in the global knowledge system. Luczaj and Holy-Luczaj refer, for instance, to the concept of Global North vs. Global South divide, metropoles vs. non-metropoles approach, and the centers–peripheries theory of Wallerstein, drawing attention to the asymmetrical spatial relations among and within countries or groups of countries (Luczaj & Holy Luczaj, 2020).

Goebel et al. organize their paper around the concept of decoloniality as a critique of Eurocentric frames of the global knowledge production system (Goebel et al., 2020). Niner and Wasserman (2021) and Goebel et al. (2020) also recall the concept or culture of "othering," pulling together diverse academic practices that exclude and marginalize non-white, non-Western, and female academics from societies, editorial boards, and international conferences. The authors speculate whether moving from F2F to virtual meetings could counteract various types of othering (Goebel et al., 2020; Niner & Wassermann, 2021).

The fifth group of concepts uncovered in VCs collection refers to the **peculiarities of technological development**. Thatcher (2006) deployed media richness theory (Daft & Lengel, 1986) and social presence theory (Short et al., 1976) to discuss the enablers and inhibitors of participation in the CybErg VCs series. Media richness theory focuses on the correspondence between the communication qualities of a particular medium and the requirements of a specific task. "Rich'' communication media allow to clarify ambiguity, provide immediate feedback, and incorporate various cues and language diversity. Social presence theory assumes that different communication media vary significantly in the degree of the psychological experience of being socially present, which depends on the number of communication channels available to convey information and cues to other social participants. Thus, F2F would be a "richer" medium and assure a higher degree of "social presence" than VCs (Thatcher, 2006).

Falk and Hagsten (2021b) used the unified theory of acceptance and use of information technology (UTAUT) (Venkatesh et al., 2003) to disclose the factors responsible for the decision to organize a virtual conference instead of cancelling or postponing the planned event. According to this theory, the use of information systems depends on performance expectancy, effort expectancy, social influence, and facilitating conditions. The first three factors relate to intended behavior, while the fourth determines usage (Falk & Hagsten, 2021b). In addition, Braun et al. (2020) recall computational turn in the social sciences and "solutionism" to stress the practices in constant need of new (online) technologies and their influence on scholarly results.

Additional numerous concepts are evoked in the selected VCs collection, including Boyer's scholarship classification system (Lessing et al., 2020), ethical theory (Parncutt & Seither-Preisler, 2019), the feminist stance (Black et al., 2020), risk assessment theory (Parncutt & Seither-Preisler, 2019), sustainability (Niner & Wassermann, 2021), and the concept of the four dimensions of academic events (Falk & Hagsten, 2021b). Four dimensions relevant for the examination of VCs are offered by Falk and Hagsten (2021b), including size, focus, participants, and tradition.

### 3.7. VC's advantages and disadvantages

The advantages and disadvantages of virtual conferences constitute one of the main topics covered by the analyzed collection – 58% of analyzed items discuss advantages or disadvantages (or both). The advantages and disadvantages of virtual conferences are discussed from diverse perspectives and distinct levels of detail. However, recurring themes and arguments can be distinguished. The advantages of virtual conferences can be divided into the following groups (ranked according to the number of publications that discuss a particular advantage):

- More inclusive and democratic scholarly communication (discussed in 67 papers)
- Costs and time savings (63 papers)
- Environmental reasons (49 papers)
- Enhancing collaboration and maintaining the scientific endeavor or professional development (42 papers)
- Flexible and tailor-made attendance possibilities (40 papers)
- Higher quality of the scholarly discussion (39 papers)
- Higher research community participation and scalability of virtual meetings (35 papers)
- Content availability (15 papers)
- Safety and health (14 papers)
- Lower organizational burden (14 papers)

Similarly, the disadvantages of virtual conferences can be divided into the following groups:

- A lack of or limited social interaction during virtual events (58 papers);
- Lesser or uneven conference engagement (42 papers),
- Overwhelming technology dependence (31 papers)
- Planning and organizational burden (26 papers)
- Security issues (26 papers)
- Lower scientific value (22 papers),
- Professional isolation and lower collaboration opportunities (21 papers),
- Endangering participants' well-being (17 papers)
- Higher requirements toward participants (12 papers)

Below, we discuss these groups in more detail.

#### 3.7.1. Advantages

**More inclusive and democratic scholarly communication**. Inclusiveness is the dominant advantage of VCs remarked in 67 items (Foramitti et al., 2021; Ha et al., 2021). Attending the virtual conference can be immersive regardless of participants' constraints (Wu et al., 2021) and their physical location (Dumova, 2012; Fang & Daniel, 2021). The analysed literature confirms that virtual conferences are more accessible than traditional events for scholars:

- with family obligations, e.g. small children or sick family members (Abbott, 2020; Achakulvisut et al., 2020; Houston, 2020; Saliba, 2020; Sarabipour, 2020), that affect woman to higher degree (Wu et al., 2021);

- under pressure of professional duties such as business responsibilities or a demanding lecturing schedule (Fellermann et al., 2020; Thatcher, 2006);

- experiencing gender (Sarabipour, 2020; D. Sharma, 2021), race and culture bias (Sá et al., 2019), including those related to religious practices (Achakulvisut et al., 2020);

- impacted by disabilities, e.g. hearing and visual disabilities (Dua et al., 2021; Fellermann et al., 2020; Lecueder & Manyari, 2000; Sarabipour, 2020), health concerns (Achakulvisut et al., 2020; Saliba, 2020; Stamelou et al., 2021; Viglione, 2020a) or chronic diseases (Weissgerber et al., 2020);

- experiencing travel bans, e.g. visa restrictions (Batra et al., 1999; Fellermann et al., 2020; Pedaste & Kasemets, 2021; Rundle et al., 2020; Saliba, 2020; Weissgerber et al., 2020);

- suffering from limited funding, which in particular affects students (Chauhan et al., 2021; Gichora et al., 2010), untenured faculty (Reinhard et al., 2020), and researchers from remote locations, inadequately funded (Fraser et al., 2017; Stamelou et al., 2021; Weissgerber et al., 2020) or representing resource-poor universities (Roos et al., 2020; D. Sharma, 2021; Viglione, 2020a)

- with language barrier (Cristia, 2019; Lecueder & Manyari, 2000; Levitis et al., 2021; Sarabipour, 2020; Thatcher, 2006);

- affected by social anxiety (Cristia, 2019; Thatcher, 2006), or with introverted personality (Ball, 2000; Busse & Kleiber, 2020; Kalia et al., 2020; Sá et al., 2019; D. Sharma, 2021);

- highly concerned about environmental issues, experiences a personal conflict between their ecological ethos and duty to travel (Fellermann et al., 2020; Forrest et al., 2020);

- representing lower or selective motivation to participate in the whole event, e.g. interested in only specific topics (Alberio, 2021; Ball, 2000).

Virtual conferences' diversity is manifested in the wider international distribution of participants (Milić et al., 2020; Thatcher et al., 2011). The reduction of conference fees and travel costs, as well as waning logistic barriers, made virtual conferences more reachable for scholars representing distant locations (Ayhan & Naqui, 2021; Bottanelli et al., 2020; Lecueder & Manyari, 2000; D. Sharma, 2021), and developing countries (Gichora et al., 2010; Sarabipour, 2020; Woolston, 2020), such as Latin America states (Forrest et al., 2020), India and China (Thatcher et al., 2011). The democratisation of attendance enables international dialogue on previously underrepresented topics of universal concern (Mubin et al., 2021; Tombaugh, 1984) and expands knowledge sharing among participants representing developed and developing countries (Fulcher et al., 2020; Roos et al., 2020).

Moreover, researchers at different stages of their careers and with various experience can be more equally represented: scientists and non-scientists, students, trainees, and laboratory technicians (Hameed et al., 2021; Salomon & Feldman, 2020). Foramitti et al. (2021) revealed that 68% of investigated VC participants valued power dynamics between participants during the event, e.g. senior vs junior researchers or dominant behaviour of some participants in debates. VCs support also researchers' possibilities to play different roles during the conference: from a passive participant, discussant, speaker, panel expert, and session chair to scientific and organising committee member (Sarabipour, 2020). Last but not least, VC makes the research results more accessible to a broader audience, practitioners, and policymakers (Moss et al., 2021; Reinhard et al., 2020; Roos et al., 2020).

The communication within diverse pool of VC participants is based on less-hierarchical rules and freed from stereotypes due to the less noticeable status cues in digital transmission (C. L. Chou et al., 2012; Green, 1998; Hanaei et al., 2020; Houston, 2020; Niner & Wassermann, 2021; Williams & Chalmers, 2015). Digital communication with eminent scholars is more direct and less intimidating than face-to-face one (Dua et al., 2021; Roos et al., 2020; Woolston, 2020). The possibility to ask questions in a written form lowers access barriers and democratises participants' involvement (Bozelos & Vogels, 2021; Houston, 2020). These processes enhance the quality and variety of conference program, (Ball, 2000; D. Sharma, 2021), and can strengthen the scientific community created during the conference and beyond (Saliba, 2020).

**Costs and time savings**. Costs and time savings resulting from the absence of travelling and physical gathering are second most evident advantages of virtual conferences stated in 63 analysed items (Busse & Kleiber, 2020; J.-Y. Chou & Camerlink, 2021; Falk & Hagsten, 2021a; Hameed et al., 2021; Wilkinson & Hemby, 2000). Most papers integrate the reduction of costs and time and adduce them among many other benefits (Achakulvisut et al., 2021; Fellermann et al., 2020; Kalia et al., 2020; O'Haver, 1995; Salomon & Feldman, 2020). Moving from a physical to a digital event decreased conference organisers' expenses (Bozelos & Vogels, 2021; Jauhiainen, 2021; Mubin et al., 2021; Niner & Wassermann, 2021).

Travelling, keynote speakers' accommodation, venue hiring and catering expenses disappear, the work overload of organisers is lesser (Schwarz et al., 2020b; Thatcher, 2006), and planned subsidies for environmentally friendly travel modes wane (Foramitti et al., 2021). The analysed collection provides numerous examples of drop off in organisers' costs: in some cases the budget was reduced by 55-60% (Viglione, 2020a), 90% (Schwarz et al., 2020b) or even 96% in comparison to the on-side events (Foramitti et al., 2021). Lower expenses of the organisers allow them to eliminate or radically reduce the registration costs depending on the participant's status (Ball, 2000; Bhargava et al., 2020; Pedaste & Kasemets, 2021).

Virtual conferences appear cost-effective for attendees because registration costs reduction (T. Anderson, 1996; D. Sharma, 2021) and no costs of travel, accommodation, and meals (Kopec & Stolbach, 2020; Parncutt & Seither-Preisler, 2019; Sipley G, 2021). Online events also allow the elimination of expenses on travel permits (Guerra Amorim & Tucci, 2020; Patel & Sobh, 2008), childcare and infant nursing facilities and attendee safety (Sarabipour, 2020). Participants perceived time savings as their opportunity benefits (Dua et al., 2021; Niner & Wassermann, 2021; Pedaste & Kasemets, 2021) resulting from the reduction of spending long hours on the journey to, from or between the conferences (Houston, 2020; Jauhiainen, 2021; Lecueder & Manyari, 2000; Raby & Madden, 2021), reduction of time-consuming bureaucracy involved in organising overseas trips (Abbott, 2020) and possibilty to fulfil other academic responsibilities at the laboratory, lecture hall or hospital (Chauhan et al., 2021; Porpiglia et al., 2020).

**Environmental reasons**. Acting on behalf of environmental sustainability is another advantage of online conferencing mentioned in 49 papers. Organising scientific events online decreases greenhouse gas emissions from flights, ground transportation, and accommodation (J.-Y. Chou & Camerlink, 2021; Donlon, 2021; Kalia et al., 2020; Lester, 2007; Rissman & Jacobs, 2020; Sá et al., 2019; Saliba, 2020; D. Sharma, 2021) and reduces unnecessary consumption, e.g. food, plastic, and paper wastes (Jauhiainen, 2021; Schwarz et al., 2020b). Several authors provide the ecological footprint size of specific conferences (Roos et al., 2020; Sarabipour, 2020; D. Sharma, 2021) and calculate the reduction connected to onlinification of conferences (Bozelos & Vogels, 2021; Moss et al., 2021). The authors spotlighted that although lowering the carbon footprint created by scholarly gatherings bothered scholars earlier, a lockdown pressed by the COVID-19 pandemic accelerated the changes in scholarly communication (Viglione, 2020a). The selected papers offer virtual conferencing as a sustainable alternative also for future international conferences (Busse & Kleiber, 2020; Fang & Daniel, 2021; Le et al., 2020b; Sipley G, 2021) and underline its long-lasting positive climate impact (Achakulvisut et al., 2021; Braun et al., 2020; Fellermann et al., 2020).

**Enhancing collaboration and maintaining scientific endeavour/professional development**. The authors of 42 papers also noticed that virtual conferences might provide much broader opportunities for scholarly communication and networking (Houston, 2020; Jauhiainen, 2021; Price, 2020; Thatcher et al., 2011; Wu et al., 2021), might help to establish novel long-term relationships (Dunn et al., 2021; Ekstrom et al., 2020; Parncutt & Seither-Preisler, 2019; Sarabipour, 2020) or expand the impact the given scholarly community (Chauhan et al., 2021; Fang & Daniel, 2021). It is feasible because virtual conferences are featured with innovative technological tools to enhance interactivity and potential collaborations. Among many examples authors listed: virtual platforms (Lessing et al., 2020; Patel & Sobh, 2008), forums (Arnal et al., 2020), online archives (Ball, 2000; Moss et al., 2021), shared virtual reality spaces (Foramitti et al., 2021; Kirchner & Nordin Forsberg, 2021; Lawrence et al., 2000), gamification tools (Leimeister et al., 2021), matchmaking apps including scientists' biographies (Viglione, 2020b), data sharing apps (D'Anna et al., 2020), and so-called introduction channels for providing essential information on participants interests (Foramitti et al., 2021). As a result the authors see the VCs as viable solution of maintaining (Porpiglia et al., 2020; Wu et al., 2021) or increasing scientific productivity (Achakulvisut et al., 2020; Hampton et al., 2017). Some regard online conferencing as a tool for improvement of transparency and reproducibility in science, and more efficient relocation of R&D funds (Sarabipour, 2020).

**Flexible and tailor-made attendance possibilities**. Another benefit revealed in 40 papers is convenience, flexibility and tailor-made attendance possibilities. VCs allow delegates to participate in the conference eliminate travel permit expenses (Bottanelli et al., 2020; Gao et al., 2020; G. Sharma & Schroeder, 2013), from wherever they can feel comfortable (Blackman et al., 2020; Patel & Sobh, 2008). The online format allows not only to combine attending the event with other duties related to private or work commitments (Goebel et al., 2020; Houston, 2020; Misa et al., 2020; Saliba, 2020), but also personalise coffee or lunch breaks (G. Sharma & Schroeder, 2013). Participants can switch among parallel sessions (Arnal et al., 2020; Castelvecchi, 2020; Dumova, 2012; Larus et al., 2021; Patel & Sobh, 2008; Thatcher, 2006; Wu et al., 2021) or selectively attend several different events during the same day (Woolston, 2020). Due to the availability of recorded talks, delegates might watch them on a personal schedule (Larus et al., 2021). That allows them to overcome the challenges of time zones and connectivity (Goebel et al., 2020; Kalia et al., 2020; Moss et al., 2021; Sá et al., 2019), to spend more time understanding the content, or to discuss it with the team members (Alberio, 2021). Virtual conferences are more fluid, as the organisers might attribute and change the roles of participants during the event (Saliba, 2020), and participants might extend or squeeze their participation time (Houston, 2020).

**Higher quality of the scholarly discussion**. 37 papers highlighted that VCs ensure more thoughtful and in-depth discussions than face-to-face meetings, particularly: higher quality of comments and questions (Busse & Kleiber, 2020; Cristia, 2019; Foramitti et al., 2021; Milić et al., 2020; O'Haver, 1995; Saliba, 2020), the superiority of submitted papers (Lecueder & Manyari, 2000; Thatcher, 2006; Thatcher et al., 2011), and better quality of presentations and poster sessions (Alberio, 2021; Fellermann et al., 2020). The higher scientific quality is due to the VCs capacity to attract high-impact scholars in the field (Chauhan et al., 2021; Gichora et al., 2010; Kalia et al., 2020; Wilkinson & Hemby, 2000), and increased diversity of participants (Ball, 2000; Chauhan et al., 2021). Moreover the authors pointed out to the wide range of factors enhancing understanding of talks and better quality of communication during the event: more time for submitting questions and discussions (Lecueder & Manyari, 2000); the straightforwardness in submitting questions for all participants (Ball, 2000; Houston, 2020; Saliba, 2020; Tombaugh, 1984; Viglione, 2020b), more incentives for moderators work (Ha et al., 2021; Price, 2020; D. Sharma, 2021); possibilities to extend the talks beyond the meeting (Achakulvisut et al., 2020; Arnal et al., 2020; Foramitti et al., 2021; Lawrence et al., 2000; D. Sharma, 2021), and better possibilities to see and hear the speakers (Achakulvisut et al., 2020; Viglione, 2020a). As a result, VCs widen the range of the scientific ideas presented (Mubin et al., 2021), enhance knowledge transfer (T. Anderson, 1996; Ekstrom et al., 2020; Hameed et al., 2021), and bring new perspectives to the scholarly conversation (Lawrence et al., 2000).

**Higher research community participation and scalability of virtual meetings**. The digital nature of scholarly meetings permits broader participation of the research community, which was mentioned in 42 papers (Ball, 2000; Donlon, 2021; Goebel et al., 2020; Green, 1998; Guerra Amorim & Tucci, 2020; Ha et al., 2021; Meyer et al., 2021; Milić et al., 2020; Mubin et al., 2021; O'Haver, 1995; Price, 2020, p. 20; Rundle et al., 2020; Speirs, 2020; Wu et al., 2021). Mikhridinova et al. (Mikhridinova et al., 2021) reported an increase in (online) delegates over conventional editions of the conference in years gone by 10% and Sarabipour et al. (Sarabipour, 2020) by 50%.

Larus et al. (Larus et al., 2021) noticed that the number of conference attendees had doubled, while Woolston (Woolston, 2020) reported four times higher registration after going digital. Moreover, (Larus et al., 2021) revealed that 67% of registrants had never attended the ASPLOS conference before moving it online, and only 31% had attended 1-5 previous ASPLOS conferences.

Authors draw attention to the scalability of online conferences. It means independence of physical space or time constraints, e.g. concerning room requirements such as volume or number of seats (Lecueder & Manyari, 2000; Patel & Sobh, 2008), of working hours of conference centres or publishers (Dumova, 2012), as well as no need to increase costs accordingly to the number of registrants. The flexible accommodation to the number of registrants allows conference organisers to respond to scholars' requests at short notice and adjust the event accordingly, e.g. address more up-to-date topics by organising additional workshops or panels (Lecueder & Manyari, 2000; Leimeister et al., 2021). Consequently, the conferences can be inclusive (Achakulvisut et al., 2020), and the engagement of participants is high (Wu et al., 2021).

**Content availability**. The significant advantage of VCs glimpsed in 14 papers is their durability, understood as long-term open access availability of the content created by the attending scholars (Achakulvisut et al., 2021; Gichora et al., 2010). The recorded presentations, abstracts, posters, discussants' questions and comments, and other research materials, such as links to additional sources or data, are freely available for participants before and after the event (Ball, 2000; Fellermann et al., 2020; Green, 1998; Ha et al., 2021; Sarabipour, 2020), often in an integrated, structured, browsable format (Lawrence et al., 2000; Sarabipour, 2020). It opens opportunities for archiving the conference results (Baird & Borer, 1987; Ball, 2000) and disseminating conference products in an open-access formula for a wider audience (Bonifati et al., 2020; Fellermann et al., 2020; Kalia et al., 2020).

**Safety and health**. As mentioned in 14 papers, attending VCs might increase participants' safety and well-being. Online format mitigates the risk of contracting COVID−19 and other infections (D'Anna et al., 2020; Dua et al., 2021; Fang & Daniel, 2021; Reinhard et al., 2020; Rekawek et al., 2020). Moreover, lack of travelling eliminates challenges related to gruelling journeys, e.g. jetlag problems (Houston, 2020; Saliba, 2020). Participants of VCs appreciate the relief of stress associated with travelling, presentation in person, and bypassing family and work commitments (Busse & Kleiber, 2020; J.-Y. Chou & Camerlink, 2021; Gao et al., 2020; Kalia et al., 2020). Attending an online format disallows inappropriate personal interactions, e.g. sexual harassment that might occur at academic conferences (Moss et al., 2021).

**Lower organisational burden**. The authors of 13 papers noticed that organising a virtual conference is logistically simpler (and speedier) than a live conference (Fraser et al., 2017; Rekawek et al., 2020). It needs a team of fewer people (Rekawek et al., 2020) assisted by a large pool of volunteers accessible online if needed (Achakulvisut et al., 2020). Key speakers might be more eager to accept the invitation because of increased scheduling flexibility compared to in-person meetings (Chauhan et al., 2021). The organisation of the online event does not require physical space booking, organising social events, and using conference infrastructures such as projectors or electronic screens (Ball, 2000; Misa et al., 2020; D. Sharma, 2021). At the same time, tracking the information on participants or surveying their feedback is much easier, and the scope of available information broader (Achakulvisut et al., 2020; Rekawek et al., 2020; Sarabipour, 2020).

**Other advantages** of VCs mentioned in the analysed collection are those related to the scholarly associations' benefits connected to lower costs and expansion of their reach (Chauhan et al., 2021; Falk & Hagsten, 2021a; Sarani et al., 2020).

### 3.7.2. Disadvantages

**The lack of or limited social interaction during virtual events**. The issue is considered from several angles, analysing the reasons, the effects and the most affected groups. The little social interaction during the conference stems from the fact that the virtual conferences' relations occur with the technological interface, which limits the detection of non-verbal clues and precludes reading body language and interpreting eye contact, which is vital for human communication (Hanaei et al., 2020; Kirchner & Nordin Forsberg, 2021; Price, 2020; Tombaugh, 1984; Williams & Chalmers, 2015). Some information is impossible to transmit through the internet (Ball, 2000). Online conference platforms can hardly reproduce empathy, human contact (Ruiz-Barrera et al., 2021), affection, and emotions (Porpiglia et al., 2020). The deficiency of social interactions is particularly detrimental for early career researchers with less established professional links (Milić et al., 2020; Saliba, 2020; Sarani et al., 2020), the extroverts (Moss et al., 2021), and interdisciplinary communities (Schwarz et al., 2020b) and participants of small, focused events (Lester, 2007).

**Lesser or uneven conference engagement**. A substantial part of the collection represented by 42 papers relates to the lesser or uneven engagement of the speakers and the audience in VC than face-to-face events (Gottlieb et al., 2020; Wu et al., 2021). The essential factor of the lesser engagement is varied access to technology and infrastructure, internet bandwidth included (Ball, 2000; Donlon, 2021; Niner & Wassermann, 2021). Moreover, the technical interface and its facets challenge presenting online (Brown, 2020), establishing a communication channel with the audience and capturing its reactions (Fang & Daniel, 2021). The lack of a captive audience (Green, 1998) and the feeling of isolation lowers emotions (Kalia et al., 2020) and flattens the tonality and body language of the speaker (Gottlieb et al., 2020). As a result, engagement and comprehension might be lower (Cristia, 2019), mainly when organisers display slides rather than a speaker's face during the lecture (Kalia et al., 2020). Open scientific debate is challenging due to the lack of pressure to listen prompted by eye contact with the presenter and the absence of other delegates (Hanaei et al., 2020; Kalia et al., 2020).

Attending virtually from home or the office is associated with the influence of various distractor elements such as activities or duties related to the family or professional life (Fraser et al., 2017; Misa et al., 2020; Salomon & Feldman, 2020; Sarani et al., 2020). Participating in the conference as a part of a daily routine impedes the total immersion experience (Dua et al., 2021; Sá et al., 2019) and creates time constraints problem (T. Anderson, 1996). Moreover, VCs presenters, attendants and discussants (Donlon, 2021) struggle with time zone differences (Dua et al., 2021; Forrest et al., 2020), which reduces their ability to make an exciting speech, hampers the sustainability of concentration during the night (Misa et al., 2020) and forge unsatisfying real-time interaction (Fellermann et al., 2020; Fraser et al., 2017). Among other less meaningful factors lowering the conference's commitment, the authors considered: a refusal of the novel VC phenomenon (Fang & Daniel, 2021), low interest in the remote tiresome presentations or poster sessions (Achakulvisut et al., 2021; Brown, 2020; Fellermann et al., 2020; Viglione, 2020a), reluctance to participate in VC debate-oriented sessions (Fellermann et al., 2020) or general lack of excitement of additional online activity (Roos et al., 2020).

The groups that might be in particular affected by the lower engagement are:

- attendees with caregiving (Levitis et al., 2021), particularly women (Braun et al., 2020; Meyer et al., 2021);
- attendees from localities with poor internet connections (Fraser et al., 2017; Ha et al., 2021);
- participants from countries with geopolitical restrictions might also not have access to required technical solutions, e.g., YouTube is not accessible from China (Levitis et al., 2021);
- young, less-experienced, and less-known scholars suffer from higher visibility to established researchers during VCs (Donlon, 2021; Hameed et al., 2021; Hanaei et al., 2020; Houston, 2020; Levitis et al., 2021);
- participants with low digital literacy (Jauhiainen, 2021; Roos et al., 2020; Sarani et al., 2020) or the technology do not assist their disabilities (Fulcher et al., 2020; Niner & Wassermann, 2021);
- scholars suffering from limited funding (Fraser et al., 2017; Weissgerber et al., 2020).

**Overwhelming technology dependence**. Vital concerns enclosed in 31 papers surround the unreliability of VC technology solutions, usually termed technology malfunctions or glitches (Achakulvisut et al., 2021; J.-Y. Chou & Camerlink, 2021; Rekawek et al., 2020; Sá et al., 2019; Woolston, 2020). The most frequently indicated failure relates to the low quality of the internet connection (Ruiz-Barrera et al., 2021; Sarani et al., 2020; G. Sharma & Schroeder, 2013).

The authors criticized sound issues related to low audio quality (Kirchner & Nordin Forsberg, 2021; Larus et al., 2021; G. Sharma & Schroeder, 2013; Woolston, 2020) or background noises (Hanaei et al., 2020; Williams & Chalmers, 2015) and to poor video quality, though less frequently (Ha et al., 2021). The discussion pointed to other issues linked to insufficient server capacity (Houston, 2020), such as low software platform efficiency (Fellermann et al., 2020) or cumbersome attributes of advanced hardware such as heavy VR headsets (Kirchner & Nordin Forsberg, 2021). Authors assessing the technical side of early-stage VC point out HTML limitations (impossibility of including equations and other complex elements) and related software limitations (Ball, 2000), low access to telecommunication terminals and costs of connection (Baird & Borer, 1987), unfriendly conference systems (Baird & Borer, 1987) and low-quality internet bandwidth which led to the necessity of splitting papers into smaller parts (Bachrach, 1995).

**Planning and organizational burden**. The selected papers often recognized the planning and development burden as an essential drawback of VC (Goebel et al., 2020; Wilkinson & Hemby, 2000). The effects of inadequate planning might be destructive to the event, as aptly summarized by Williams and Chalmers (Williams & Chalmers, 2015): “Non-engaged participants typing e-mails, loud background noises and failure of the chair to bring in all participants.”

Since the beginning of VCs history, the organizers and participants have struggled with various challenges. Some of the issues, such as blocking telephone lines to sustain participation (Tombaugh, 1984) or lacking the possibility of simultaneous translation (Green, 1998), have been eliminated by technological development. Nevertheless, challenges remain. For example, scheduling and coordinating VCs are logistically tricky. Development of a meticulous and easily navigated program with transparent login instructions demands due effort from the organizers. They also need to be prepared for altering levels of participants’ technical skills, technological infrastructure (Sarani et al., 2020; Tombaugh, 1984), and differentiated personal commitments (Mubin et al., 2021). Some new organizational predicaments occurred. Time zones present the most significant obstacle for simultaneous online events with a globally distributed audience (Fraser et al., 2017; Larus et al., 2021; Levitis et al., 2021; Meyer et al., 2021; Moss et al., 2021; Mubin et al., 2021; D. Sharma, 2021). Securing just-in-time technical help (Sarani et al., 2020) or the assurance of smooth communications during the sessions (G. Sharma & Schroeder, 2013) requires additional backstage work before and during the conference (Bilas et al., 2020), such as detailed planning (Bonifati et al., 2020) and the training of volunteers (Cristia, 2019).

Forced transition to the online format is exceptionally challenging, particularly when conference organizers simultaneously confront the logistical and financial impact of canceling an F2F event (Houston, 2020; Weissgerber et al., 2020; Woolston, 2020).

A fair estimation of VC costs is also problematic. Pre-COVID-19 papers estimated costs of conducting a virtual conference as relatively low; however, the organizational burden was noted (Ball, 2000; Cristia, 2019). Organizational risk of converted events stems from two contradicting phenomena: the uncertainty of the number of participants and high reduced conference fee expectations (Pedaste & Kasemets, 2021).

**Security issues**. Misgivings of confidentiality and security issues in online communication was considered in 26 papers (Jauhiainen, 2021; Mubin et al., 2021). Authors refer in particular to hackers attacks or so-called zoom-bombing – unethical practices that devastate VC sessions by broadcasting damaging and abusive content (Bonifati et al., 2020; Brown, 2020; Donlon, 2021; Goebel et al., 2020; Luczaj & Holy Luczaj, 2020; Ruiz-Barrera et al., 2021; Schwarz et al., 2020b). Pre-video virtual conferences suffered from dubious reliability of websites, posts and mailing lists (Baird & Borer, 1987; Ball, 2000; Batra et al., 1999; Lecueder & Manyari, 2000). For instance, Tombaugh (Tombaugh, 1984) revealed that 61% of the scientists expressed concern about the accuracy of the information.

Also, the uncertainty related to intellectual property rights, mainly unpublished works (Ball, 2000; Bottanelli et al., 2020; Sá et al., 2019; Tombaugh, 1984) and data protection accompanies virtual conferences (Levitis et al., 2021; Schwarz et al., 2020b). Significantly, the risk is not limited to research data but also personal data embodied in unauthorized screenshots or photographs (Bottanelli et al., 2020).

On the other hand, though, security measures undertaken by the organizers and technology providers may hinder the ability to join and reliably participate in the event. It may affect less familiar or poorly equipped delegates (Fulcher et al., 2020; Sarani et al., 2020).

**Lower scientific value**. According to Niner & Wassermann (2021), before 2020 resistance to moving events online resulted from the perception that the value of conferences cannot be cultivated in this mode. The theme of the value recurs throughout multiple publications and involves interrelated epistemic, social, and technical aspects of virtual conferences.

Concerns about the content of the conference such as quantity of information, low quality information and lack of substantial contributions resulting from lack of quality control are mentioned in the earliest works (Baird &

Borer, 1987; Tombaugh, 1984) and persist through the years (Bachrach, 1995; Hameed et al., 2021; Hanaei et al., 2020; Lawrence et al., 2000). As noted by Tombaugh (1984), early VC participants from the industrialized countries were more concerned about the quality and quantity of information than about technical issues. (Hameed et al., 2021) voice concerns that online events available to the public might lack appropriate vetting of the content, disclosure of conflicts of interests, and format for learning. Finally, Lecueder and Mayari (2000) indicated general doubts about the internet as an easily available information resource, but lacking quality control. Lawrence et al. (2000) considered attracting good quality papers and maintaining a good quality discussion the main challenge of VC.

Looking from the speakers' perspective, lack of response to contributions is a problem (Tombaugh, 1984), because the reaction from the community cannot be observable (Sá et al., 2019), which complicates assessing how the talk was received (Roos et al., 2020). Moreover, virtual format might prevent speakers from engaging sufficiently with the audience (Ha et al., 2021) and hinder requesting feedback, especially in the asynchronous mode (Busse & Kleiber, 2020).

According to Brown (Brown, 2020), conferencing online may pose challenges for the chance of dispute and new ideas emerging through interaction. Similarly, Jauhiainen (2021) observes that limited networking and interaction hamper development relations for novel ideas and, eventually, innovations.

**Professional isolation and lower collaboration oppor**tunities. The VCs fail to replicate opportunities for informal, incidental contacts (Fraser et al., 2017; Niner & Wassermann, 2021; Roos et al., 2020). Compared to face-to-face, VCs are short of conventional incentives and informal spaces for networking, such as food gatherings, sightseeing, and tourist events (Busse & Kleiber, 2020; Kopec & Stolbach, 2020; Lecueder & Manyari, 2000). The technology-dependent scholarly communication results in a lack of impromptu meetings, random hallway chats or other situations conducive to serendipity (Bozelos & Vogels, 2021; Houston, 2020; Kalia et al., 2020; Meyer et al., 2021; Moss et al., 2021; Price, 2020; D. Sharma, 2021; Viglione, 2020b). Bozelos and Vogels (Bozelos & Vogels, 2021) pointed out the importance of the so-called night science language, which is a "fast and informal, stripped-down counterpart to the rigorous 'day science' language of a public presentation or a scientific publication" arguing that it also influences the depth of the scholarly discussion. Some delegates evaluate virtual meetings as lost opportunities to meet and chat with peers (Lester, 2007), and effective scientific teambuilding (Kalia et al., 2020), compromising most fundamental by-products of physical gatherings, socialization and networking in virtual events (Hanaei et al., 2020).

The limited social interactions rooted in the lack of trust among delegates (Bozelos & Vogels, 2021; Niner & Wassermann, 2021; Saliba, 2020) impact VC participants in complex and multilevel ways. Firstly, it hampers establishing group synergy and impairs socialization processes (Ekstrom et al., 2020; Green, 1998; Hanaei et al., 2020). The networking activities are limited (Ayhan & Naqui, 2021; Mubin et al., 2021; Pedaste & Kasemets, 2021), the knowledge flows on new initiatives, papers and funding are confined (Sá et al., 2019) and connecting with new people, employees or employers are blocked (Reinhard et al., 2020). The lack of informal scientific discussions in the form of feedback and comments engenders professional isolation (Brown, 2020; Schwarz et al., 2020b; Wilkinson & Hemby, 2000). Consequently, collaboration opportunities during VCs and the possibilities of long-term sustainable teambuilding after the event are weaker than during F2F scholarly meetings (Porpiglia et al., 2020; Woolston, 2020). Lack of space for informal interaction (Salomon & Feldman, 2020), psychological distance unfavorable to generating new ideas (Lawrence et al., 2000), and insufficient intercultural exchanges empowering better understanding of the nuances of collaborative research (Hanaei et al., 2020) are seen as drawbacks of VCs in that respect.

**Endanger for participants' well-being**. A remarkable number of papers (17) focus on the harmful influence of the intensive use of digital tools on VC participants' well-being, happiness, and mental health. The authors draw attention to lockdown-related social life limitations, higher stress due to loneliness and blurring boundaries between work and home, loose human contact, affection and emotions (Ha et al., 2021; Schwarz et al., 2020b). Young scholars were affected the most because of their lifestyle and limited support from their more experienced colleagues in guiding them through the conference (Levitis et al., 2021).

The direct transmission of face-to-face conferences into online format resulted in long hours spent at the screen and energy draining caused by sustaining attentiveness (Brown, 2020; Gottlieb et al., 2020; Meyer et al., 2021). Scholars suffer from the well-described experience of teleworkers described as "the sense of tiredness that can arise from continuous or overuse of virtual video-conferencing platforms (Donlon, 2021), known under several terms: "online fatigue" (Saliba, 2020), "zoom fatigue" (Busse & Kleiber, 2020; Meyer et al., 2021) or "digital-meeting fatigue" (Moss et al., 2021), "screen fatigue" (Foramitti et al., 2021) or "digital burnout" syndrome (Ayhan & Naqui, 2021; Hameed et al., 2021).

The participants' well-being was also negatively affected by time zone differences (Misa et al., 2020) and by concentrating on the technical issues, particularly maintaining a stable internet connection, during the presentation (Gao et al., 2020).

**Higher requirements toward participants**. In thirteen papers, higher requirements toward participants of VCs were indicated as the weak point. The online conference creates new technical requirements for attendees, and the preparation for the virtual format requires both costs and skills (Levitis et al., 2021; D. Sharma, 2021; Wilkinson & Hemby, 2000). Moreover, work overload before and during the conference might act discouragingly on participants. In the case of email-based events, this problem concerned the most prominent conference members, who were expected to answer countless e-mails with little reward (Tombaugh, 1984). Green (Green, 1998) pointed out that an adequate level of written literacy was also a must. Nowadays, preparatory work such as reading or watching pre-recorded speeches, is needed for full engagement in the conference debates (Fang & Daniel, 2021).

**Less commonly covered CV's disadvantages comprised**: science societies' financial problems, degradation of communication norms, carbon footprint, and elimination of conference intangibles. First, the virtualization of scientific events forced scientific societies to adapt their financial model to the shrinkage in their core revenue streams (Dua et al., 2021; Goebel et al., 2020; Kalia et al., 2020). Although the cost of organizing VCs is lower, setting the conference fee is burdened by the high level of uncertainty (Falk & Hagsten, 2021a). Second, several authors identified the issue of blurred cues and norms in online communication and reported on toxic participants' behavior or degradation of politeness, particularly in typing comments to the presentations (Baird & Borer, 1987; Foramitti et al., 2021; Kalia et al., 2020; Niner et al., 2020; Niner & Wassermann, 2021). Third, carbon emissions associated with online data traffic and the need to include them in estimating the environmental impact of online events were recognized (Braun et al., 2020; Donlon, 2021; Niner et al., 2020). Fourth, the virtual format reduces the allure of awards ceremonies (D. Sharma, 2021), and eliminates the positive feeling of being a member of the scientific family by attending the conference (Kalia et al., 2020) or experiencing a round of applause at the end of the talk (Brown, 2020). Finally, a few authors tackled long-lasting consequences for the scientific system, which the above deficiencies may force: (a) low research progress as a consequence of a lack of scientific discussion and incentives for scholarly cooperation (Milić et al., 2020); (b) dependency from the policies of large technology providers that might be in clash with scientific freedom (Braun et al., 2020; Schwarz et al., 2020b); (c) growing competition among different modes of conferences (Ball, 2000; Forrest et al., 2020), and (d) disclosure of the new forms of exclusions stemming from the pressure regarding the use of technical solutions and execution of various forms of professional (and private) activities simultaneously (Braun et al., 2020).

## 4. Discussion

The VCs phenomenon is highly dynamic, which means that technologies and technical solutions applied for the VCs' needs are becoming increasingly sophisticated; likewise, the expectations for social interaction reinforcement during VCs are also growing (Hampton et al., 2017). The early papers on VCs recommended introducing video and voice technologies to boost social interactions during virtual events (Bachrach, 1995). Nowadays, the authors draw attention to the necessity of the introduction of new solutions such as virtual and augmented reality, gamification, interactive sessions, and avatar-based video (Bonifati et al., 2020; J.-Y. Chou & Camerlink, 2021; Stamelou et al., 2021), boosting motivation for new forms of networking among participants (Levitis et al., 2021), complement participants' skills and their habits' change (Lawrence et al., 2000).

The advantages and disadvantages of virtual conferences stem from two factors. The first is the lack of travel and gathering, and the second is the intermediation of technology interface in scholarly communication. These two circumstances cause a series of interconnected positive and negative effects on academic life. The analysis of the advantages and disadvantages of VC' revealed that the application of technologies to scholarly communication affects conference participants likewise a double-edged knife. Developing interactive-friendly technologies increases the immersiveness of the conference experience. However, this process simultaneously induces new barriers to participation and deepens the exclusiveness of particular groups.

Moreover, the organizational burden related to the conferences might be seen as lower or as higher than traditional meetings depending on the organizer, the meeting character, expectations, the level of technological advancement and skills of delegates. A dependence on technology might be discouraging for less equipped participants, yet it simultaneously makes the conference results more sustainable and adds flexibility to attendance possibilities. As a safe alternative for F2F meetings in times of a pandemic, VCs meet the critiques of those recognizing the endangerment of participants' well-being related to technology fatigue. An intensive discussion considers the possibilities of collaboration as well as the scientific value of VC output. Although supporters and opponents presented their arguments, they are not reinforced by in-depth research results. Consequently, the overall picture is blurry.

The advantages and disadvantages of VC are discussed from different overlapping perspectives. Firstly, it is clear that we can differentiate between individual versus organizational perspectives on VCs pros and cons. Moreover, these perspectives can be further specified according to the individual's roles during the conference, or different demographic characteristics. Likewise, the organizational perspective is different in the case of a scholarly association, university or publisher. The second distinction relates to the perspectives of different types of conferences or adopted definitions. For instance, the opportunities for social networking were identified as higher in the face-to-face setup than in hybrid and online formats. Contrary, the reach of the audience was found to be higher in hybrid and face-to-face conferences than in webinars (Hameed et al., 2021).

Based on insights from the debate, we expect that the online component will remain a vital element of scholarly communication. However virtual conferences became increasingly important in the digital age, 'offline' meetings will continue to resonate in the field of scientific endeavor. The pandemic experience served as a forced laboratory, highlighting that it is logistically feasible to organize and hold conferences completely online. Considering the pros and cons of various conferencing modes, the question already on the agenda of acknowledged scientific associations regards new intermediate forms of scholarly gathering, such as alternating in-person and virtual conferences. Broader incorporation of hybrid forms of scholarly communication might lead to negative effects of participants' stratification depending on the type of involvement or to new forms of exclusion. Opening traditional conferences for virtual participation might introduce a two-class scientific society where virtual attendees (who are unable or unwilling to fly more, presumably mainly early career scientists) are disadvantaged compared to onsite participants (Fellermann et al., 2020). Thus, a research agenda for various and sometimes competing forms of scientific conferences ought to be developed and realized. A useful framework to identify and consider the dilemmas involved in going online represents the responsible research and innovation approach (Braun et al., 2020).

## 5. Conclusion: Toward a virtual conference research agenda

Our review has demonstrated that the theoretical layer of the research related to VCs is thin. In the literature, authors have displayed a fragmented use of selected theories to describe the paper's general context. Other than a few examples, selected papers rarely mention a theoretically based empirical model or framework and hardly ever discuss the research findings within a theoretical framework. It is exceedingly rare for a work to present a theoretical framework specifically designed for understanding scientific conferences (e.g., Hansen et al., 2020). Consequently, the existing literature on VCs is largely descriptive in nature without a clear increase in knowledge or understanding of the phenomenon. Research on VCs in science is still at an early exploratory stage. Thus, the creation and dissemination of a solid theoretical framework for studying the phenomenon is a crucial task of contemporary science.

The outcomes of this review stress the need for further research to better comprehend how VCs are described and defined. In addition, studies are needed regarding the weak and strong points of VCs, which theoretical concepts might be employed in VC analysis, and what data and methods can be used to investigate them. We showed that studies on the functions, development, and impact of conferences in science studies are weakly developed; however, they combine various research areas (González-Santos & Dimond, 2015). Therefore, we advocate establishing a distinct interdisciplinary scientific conference research subfield. The subfield should encompass the development of a theoretical framework (or competing frameworks) that will guide empirical studies. Based on our review, we recommend that future research on scholarly VCs should focus in particular on:

- the effects of VCs on scientific collaboration, productivity, and quality.

- advantages and disadvantages from the point of view of various groups of scholars to understand how conferences are utilized by particular groups of individuals (conference roles, demographic characteristics, stage of career, factors, and the level of vulnerability to exclusion) and whether VCs magnify or diminish existing divisions.

- the impact of VCs on broadening and diversifying participation in scholarly communication, in particular, whether VCs are broadening participation for previously excluded scholars.

- the effect of VC on science in developing countries; comparison of the functioning and effects of VCs in countries of the Global North and Global South.

- the environmental impact of virtual vs F2F conferences.

**Contributorship statement**

Conceptualization, Formal analysis, Investigation, Methodology, Writing – original draft, Writing – review & editing: Agnieszka Olechnicka, Adam Ploszaj, Ewa Zegler-Poleska

Data curation: Ewa Zegler-Poleska

Funding acquisition: Adam Ploszaj

**Funding**

This work was supported by the National Science Centre Poland under Grant 2018/31/B/HS4/03997.

**Data statement**

We have offered the entire set of publication metadata online as a Zotero Group “Virtual Academic Conferencing 1984–2021”, available at https://www.zotero.org/groups/4832623/virtual_academic_conferencing_1984-2021.